\documentclass{IEEEoj}
\usepackage{cite}
\usepackage{amsmath,amssymb,amsfonts}
\usepackage[nolist]{acronym}
\usepackage{algorithm}
\usepackage{algpseudocode}
\usepackage{array}
\usepackage{booktabs}
\usepackage{graphicx,color}
\usepackage{textcomp}
\def\BibTeX{{\rm B\kern-.05em{\sc i\kern-.025em b}\kern-.08em
    T\kern-.1667em\lower.7ex\hbox{E}\kern-.125emX}}
\AtBeginDocument{\definecolor{ojcolor}{cmyk}{0.93,0.59,0.15,0.02}}
\def\OJlogo{\vspace{-4pt}\hskip-4pt\includegraphics[height=18pt]{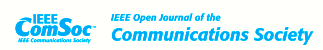}}

\DeclareMathOperator{\trace}{tr}

\newcommand{\commenttxtR}[1]{{#1}}
\newcommand{\commenttxt}[1]{{#1}}

\begin{document}
\begin{acronym}
        \acro{3GPP}{3rd generation partnership project}
        \acro{5G}{fifth generation of mobile networks}
        \acro{6G}{sixth generation of mobile networks}
        \acro{AO}{alternating optimization}
        \acro{AWGN}{additive white Gaussian noise}
        \acro{BS}{base station}
        \acro{BD-RIS}{beyond diagonal reconfigurable intelligent surface}
        \acro{CSI}{channel state information}
        \acro{DFT}{discrete Fourier transform}
        \acro{ETSI}{European Telecommunications Standards Institute}
        \acro{LOS}{line-of-sight}
        \acro{MIMO}{multiple-input multiple-output}
        \acro{MU-MIMO}{multi-user multiple-input multiple-output}
        \acro{MU-MISO}{multi-user multiple-input single-output}
        \acro{MLP}{multilayer perceptron}
        \acro{NLOS}{non-line-of-sight}
        \acro{NN}{neural network}
        \acro{OFDM}{orthogonal frequency division multiplex}
        \acro{ReLU}{rectified linear unit}
        \acro{RIS}{reconfigurable intelligent surface}
        \acro{SCA}{successive convex approximation}
        \acro{SD}{sphere decoding}
        \acro{SISO}{single-input single-output}
        \acro{SIM}{stacked intelligent metasurface}
        \acro{SINR}{Signal-to-interference-plus-noise ratio}
        \acro{SNR}{signal-to-noise ratio}
        \acro{STM}{strongest tap maximization}
        \acro{SVD}{singular value decomposition}
        \acro{UE}{user equipment}
\end{acronym}
\receiveddate{XX Month, XXXX}
\reviseddate{XX Month, XXXX}
\accepteddate{XX Month, XXXX}
\publisheddate{XX Month, XXXX}
\currentdate{11 January, 2024}
\doiinfo{OJCOMS.2024.011100}

\title{Tree Search Algorithms Applied to the BD-RIS Configuration in MU-MISO Communication Systems}

\author{PEDRO~H.~C.~DE~SOUZA\IEEEauthorrefmark{1} AND LUCIANO~MENDES\IEEEauthorrefmark{1} \IEEEmembership{(Member, IEEE)}}
\affil{National Institute of Telecommunications - Inatel, Sta. Rita do Sapucaí, MG 37536-001 BRAZIL}
\corresp{CORRESPONDING AUTHOR: Pedro~H.~C.~de~Souza (e-mail: pedro.carneiro@dtel.inatel.br).}
\authornote{This work has been funded by the following research projects: Brasil 6G Project with support from RNP/MCTI (Grant 01245.010604/2020-14), xGMobile Project code XGM-AFCCT-2024-2-15-1 with resources from EMBRAPII/MCTI (Grant 052/2023 PPI IoT/Manufatura 4.0) and FAPEMIG (Grant PPE-00124-23), SEMEAR Project supported by FAPESP (Grant No. 22/09319-9), SAMURAI Project supported by FAPESP (Grant 20/05127-2), Ciência por Elas with resources from FAPEMIG (Grant APQ-04523-23), Fomento à Internacionalização das ICTMGs with resources from FAPEMIG (Grant APQ-05305-23), Programa de Apoio a Instalações Multiusuários with resources from FAPEMIG (Grant APQ-01558-24), Redes Estruturantes, de Pesquisa Científica ou de Desenvolvimento Tecnológico with resources from FAPEMIG (Grant RED-00194-23), and Centro de Competência em Redes 5G e 6G Ref. Finep nº 1060/2 contract nº 0-1-25-0883-00. This work has also been supported by a fellowship from CNPq and xGMobile/EMBRAPII/MCTI.}
\markboth{Preparation of Papers for IEEE OPEN JOURNALS}{Author \textit{et al.}}

\begin{abstract}
The \ac{RIS} has attracted considerable attention from both academia and industry in recent years, given its capacity to dynamically manipulate the reflection of incident electromagnetic waves. Although the research developed for the \ac{RIS} may have reached its maturity, there are still contentious aspects and limitations regarding its potential benefits for the next generation of wireless communications. In order to improve upon the \ac{RIS} technology, the \ac{BD-RIS} was recently proposed as a promising alternative. The \ac{BD-RIS} boasts a more sophisticated circuit topology that is capable of providing more combinations of different adjustments or configurations for signal reflection. However, to aptly reap the benefits of the \ac{BD-RIS}, the added degrees-of-freedom of its configuration must be leveraged accordingly. Therefore, in this work we propose a depth-first tree search algorithm for configuring the \ac{BD-RIS} in \ac{MU-MISO} communication systems. Taking advantage of the tree search exploration, the proposed algorithm achieves a remarkable trade-off between channel strength maximization performance and computational complexity scalability.
\end{abstract}


\maketitle

\section{INTRODUCTION}
\acresetall
\IEEEPARstart{I}{n} recent years the \ac{RIS} has been heralded as a potential candidate solution for integrating the ecosystem of the \ac{6G}. Notably, the \ac{RIS} is already being considered as a key item in the ongoing \ac{ETSI} standardization efforts \cite{lix:25}, for example. The \ac{RIS} consists of a planar array of tunable resonant elements designed to dynamically and precisely manipulate the reflection of incident electromagnetic waves. It can be seen as a byproduct of the research in metasurfaces, conducted in the context of signal transmission in communication systems \cite{jorge:25}. Although the \ac{RIS} is already recognized as a mature technology, in that the initial steps of research are well consolidated, some aspects of its implementation and practical limitations are still being worked out by academia and industry \cite{kara:25}.

In the face of limitations presented by \acp{RIS} systems, other alternatives are being proposed that improve upon the \ac{RIS}, \commenttxt{such as \acp{SIM}. These are stacked metasurfaces capable of native signal processing in the wave domain, with the most recent advances obtained with spherical \acp{SIM} \cite{zhang:26} being able to achieve more uniform full-space coverage and higher average spectral efficiency. Likewise, other emerging solutions based on \ac{RIS} systems are also being put forward,} one being the so-called \ac{BD-RIS} \cite{li:24,li:25}.

Similarly to the \ac{RIS}, in principle the \ac{BD-RIS} also relies on adjustable resonant elements or impedances to alter the impinging wave properties, except that for the \ac{BD-RIS} additional impedances are used to interconnect these elements. In other words, while for the \ac{RIS} the impinging signal is reflected directly by the adjusted elements, for the \ac{BD-RIS} the impinging signal flows through a more sophisticated circuit topology of interconnected impedances. This allows more combinations of different adjustments or configurations to be made available for \ac{BD-RIS} systems. Hence the term beyond diagonal, since its configuration is represented mathematically by a matrix, whereas for \ac{RIS} systems it is restricted to a diagonal matrix. It is envisioned that the \ac{BD-RIS} could be deployed in future communication systems with the aim of improving the signal transmission \commenttxtR{through} the communications channel \cite{kasta:24}. Consequently, the added degrees-of-freedom of the \ac{BD-RIS} configuration matrix can, in turn, be used to improve the overall performance in comparison with the \ac{RIS}.

By configuring the \ac{BD-RIS} accordingly, favorable propagation conditions are established for the signal transmission, this being even more impactful if the direct channel between the transmitter and receiver is weak or blocked. Therefore, we aim to configure the \ac{BD-RIS} matrix in order to maximize the overall channel strength, considering both the direct channel and the cascade channel between transmitter-\ac{BD-RIS}-receiver. It is well known that directly adjusting the \ac{BD-RIS} matrix entries (impedance values) for channel strength maximization \commenttxt{leads to} a prohibitive number of combinations comprising the search region. Consequently, the channel strength maximization can be otherwise regarded as a non-convex optimization problem \cite{fang:24,santa:26,zhang:25,fida:25}, with the \ac{BD-RIS} matrix as the optimization variable. However, this approach can still suffer from high computational complexity and limited system scalability. Alternatively, conventional and more simple heuristic search algorithms have the potential to be more accessible, presenting a satisfactory balance between channel strength maximization performance and computational complexity scalability across different \ac{BD-RIS} matrix sizes.

Search algorithms have already been used to configure the diagonal \ac{RIS} under diverse communications scenarios and with different objectives. In \cite{quispe:21}, the authors propose a Branch-Reduce-and-Bound search algorithm to find the optimal \ac{RIS} configuration in order to maximize the energy efficiency for aerial-RIS-aided cooperative wireless communication systems. Moreover, the authors of \cite{ahmed:23} propose an information-theoretic tree search for configuring the \ac{RIS} in massive \ac{MIMO} communication systems. The work \cite{xiong:24} also introduces a divide-and-sort search algorithm for maximizing the received signal power, considering the joint optimization of the transmit beamforming vector and the \ac{RIS} configuration. Alternatively, the authors of \cite{rame:25} present an algorithm inspired by the \ac{SD} \cite{vit:99} for optimizing the \ac{RIS} configuration, subsequently developing a low-complexity Schnorr-Euchner \ac{SD} algorithm that can efficiently obtain the \ac{RIS} configuration without significant loss in optimality. Finally, in \cite{lin:25} a Branch-and-Bound algorithm optimizes the global \ac{RIS} deployment to maximize overall signal gain in shadow areas, that is, wireless communication blind spots where there are insufficient signal coverage.

In this work we propose a depth-first tree search algorithm for configuring the \ac{BD-RIS} matrix. \commenttxt{Since one of the aims of our work is to present the most scalable algorithm in terms of computational complexity, we have chosen to implement a customized (greedy) depth-first search instead of a more complex breadth-search based search, for example \cite[Sec.~3.4.3]{russell:10}.} To the best of authors' knowledge, this is the first work that integrates tree search algorithms with \ac{BD-RIS} configuration matrices. \commenttxt{More specifically, the following main contributions are presented in this work:
\begin{itemize}
    \item A flexible tree search algorithm that allows an arbitrary mapping between tree levels and \ac{BD-RIS} matrix entries and computes off-diagonal entries efficiently;
    \item Normalized heuristic parameters that enable pruning in multiple stages of the search, so that a flexible trade-off between performance and complexity is attained;
    \item The unitary and symmetric constraints for the \ac{BD-RIS} matrix are appropriately incorporated into the search via successive matrix projections. 
\end{itemize}}
We \commenttxt{also} demonstrate that the proposed algorithm can reach the channel strength maximization upper bound for \ac{SISO} systems, and a \commenttxt{comparable} performance/complexity trade-off \commenttxt{to the manifold optimization method \cite{santa:26} when considering} \ac{MU-MISO} systems. \commenttxt{For this, extensive computational simulations are carried out, where the performance of the proposed algorithm is evaluated for different parameter configurations and the average runtime is also used to validate computational complexity predictions.}

The remainder of this work is organized as follows: Section~\ref{sec:sysModel} presents the \ac{MU-MISO} system model followed by the definition of the channel strength maximization problem; for Section~\ref{sec:SISO} the aforementioned system model is specialized to \ac{SISO} systems, so that a primer on the proposed algorithm is provided for ease of understanding; Section~\ref{sec:MUMISO} then introduces the proposed depth-first tree search algorithm with pruning for configuring the \ac{BD-RIS}; Section~\ref{sec:numResult} shows results obtained from computational simulations that validate the performance of the proposed algorithm, for which the computational complexity is also evaluated; Section~\ref{sec:future} summarizes the open problems and future research related to the proposed algorithm and Section~\ref{sec:concl} concludes the paper.

\subsection{Notation}
Throughout this paper, italicized letters (e.g. $x$ or $X$) represent scalars, boldfaced lowercase letters (e.g. $\mathbf{x}$) represent vectors, and boldfaced uppercase letters (e.g. $\mathbf{X}$) denote matrices. The entry on the $i$th row and $j$th column of the matrix $\mathbf{X}$ is denoted by $X_{ij}$. The number of elements contained in a set $\mathcal{X}$ is denoted by $\left\lvert\mathcal{X}\right\rvert$. The sets of vectors (matrices) of dimension $X$ ($X\times Y$) with real and complex entries are correspondingly represented by $\mathbb{R}^{X}$ ($\mathbb{R}^{X\times Y}$) and $\mathbb{C}^{X}$ ($\mathbb{C}^{X\times Y}$). The transposition and conjugate transposition operations of a vector or matrix are represented as $\left(\cdot\right)^{\top}$ and $\left(\cdot\right)^{\dagger}$, respectively. The absolute value of the scalar $x \in \mathbb{R}$ or the modulus of $x \in \mathbb{C}$ is denoted by $\lvert x\rvert$. The real and imaginary parts of $z \in \mathbb{C}$ are denoted by $\Re(z)$ and $\Im(z)$. The $\ell_p$-norm, $p \geq 1$, of the vector $\mathbf{x} \in \mathbb{C}^{N}$ is given by $\|\mathbf{x}\|_p$. The squared Frobenius norm of a matrix, $\boldsymbol{X}$, is given by $\|\boldsymbol{X}\|_F^2 = \trace{(\boldsymbol{X}^{\dagger}\boldsymbol{X})}$, wherein $\trace{(\cdot)}$ computes the sum of the diagonal entries of a matrix. The nuclear norm is represented by $\|\boldsymbol{X}\|_* = \sum_i \sigma_i(\boldsymbol{X})$, where $\sigma_i(\boldsymbol{X})$ is the $i$-th singular value of $\boldsymbol{X}$. Finally, $x \sim \mathcal{N}\left(0,1\right)$ is a random value drawn from the normal distribution.

\section{System Model}\label{sec:sysModel}
Let a \ac{MU-MISO} system be represented by a \ac{BS} with $L$ antennas serving $K$ single-antenna \acp{UE}, with the assistance of a \ac{BD-RIS} containing $N$ elements or reflectors. The \ac{BD-RIS} is assumed to be ideal, in the sense that it provides lossless signal reflection and that across its elements perfect matching is assured and mutual coupling is negligible\commenttxt{\footnote{\commenttxt{Note that hardware imperfections may have a significant impact on \ac{BD-RIS} systems, as demonstrated for \ac{RIS} systems in \cite{pedro:25_b}.}}}. Moreover, the \ac{BS} and all $K$ \acp{UE} are assumed to be located within the half-space coverage area of the \ac{BD-RIS}. \commenttxt{Therefore, the overall channel is given by the cascade channel model \cite{li:25,fang:24}}, described by
\begin{equation}\label{eq:chModel}
    h_{\text{MU-MISO}}\left(\boldsymbol{\Theta}\right) = \boldsymbol{G}^\dagger + \boldsymbol{H}^\dagger \boldsymbol{\Theta}\boldsymbol{\Upsilon}\text{,}
\end{equation}
wherein $\boldsymbol{G} = \left[\boldsymbol{g}_1\text{,}\boldsymbol{g}_2\text{,}\dots\text{,}\boldsymbol{g}_K\right]$ and $\boldsymbol{H} = \left[\boldsymbol{h}_1\text{,}\boldsymbol{h}_2\text{,}\dots\text{,}\boldsymbol{h}_K\right]$, for which $\boldsymbol{g}_k \in \mathbb{C}^L$, $k \in \left[1\text{,}2\text{,}\dots\text{,}K\right]$, is the complex-valued channel between the $k$-th \ac{UE} and the multi-antenna \ac{BS}, and $\boldsymbol{h}_k \in \mathbb{C}^N$ represents the complex channel gains between the $k$-th \ac{UE} and all elements of the \ac{BD-RIS}. Furthermore, the entries of $\boldsymbol{\Upsilon} \in \mathbb{C}^{N \times L}$ contain the complex channel gains between the \ac{BS} antennas and all \ac{BD-RIS} elements. Finally, the \ac{BD-RIS} is denoted by the phase-shift matrix $\boldsymbol{\Theta} \in \mathbb{C}^{N \times N}$, in which its entries are given by $\Theta_{ij} = e^{\jmath \theta_{ij}} \in \mathbb{C}$.

\subsection{Problem Formulation}
See in \eqref{eq:chModel} that the overall channel depends on the configuration determined for the \ac{BD-RIS} matrix. As such, typically one wishes to configure the \ac{BD-RIS} in order to maximize the signal strength received by all \acp{UE}, thereby increasing the channel capacity. In other words, the \ac{BD-RIS} provides beam steering capabilities, that is, passive beamforming, that can be used to enhance the signal transmitted by the \ac{BS}. Therefore, in this work, we focus on the channel strength maximization described by the following problem:
\begin{equation}\label{eq:maxProb}
    \begin{aligned}
        \max_{\boldsymbol{\Theta}} \ & \|\boldsymbol{G}^\dagger + \boldsymbol{H}^\dagger \boldsymbol{\Theta}\boldsymbol{\Upsilon}\|_F^2 \\
        \text{s.t. } \ & \boldsymbol{\Theta}^\dagger \boldsymbol{\Theta} = \mathbf{I},
    \end{aligned}
\end{equation}
where the unitary constraint for the \ac{BD-RIS} configuration matrix ensures that the reflected signal power is no larger than the power of the signal impinging on the \ac{BD-RIS} \cite[Sec.~III-B]{li:25}.

Although the problem \eqref{eq:maxProb} is non-convex, an optimal solution can still be provided as long as the unitary constraint is relaxed, as demonstrated in \cite{fang:24}, for example, but its computation is costly, meaning that it entails a high computational complexity. Nonetheless, more recently it has been shown \cite{santa:26,zhang:25,fida:25} that manifold optimization methods are able to optimally solve problem \eqref{eq:maxProb} without the need to relax the unitary matrix constraint. Alternatively, instead of relying on the convexity properties of such problems, we propose a heuristic tree search algorithm for configuring the \ac{BD-RIS}. In the next section, initially the proposed algorithm is introduced considering the \ac{SISO} system model, which is specialized from \eqref{eq:chModel}. This is done with the aim of providing a more clear grasp of the algorithm essentials, before presenting its final formulation for the \ac{MU-MISO} system.              

\section{\ac{BD-RIS} Configuration: \ac{SISO} System}\label{sec:SISO}

For ease of presentation, in this section we assume that the direct channel between the \ac{BS} and \ac{UE} is obstructed. Consequently, from \eqref{eq:chModel}, the overall channel for the \ac{SISO} system can be represented as
\begin{equation}\label{eq:chModelSISO}
    h_{\text{SISO}}\left(\boldsymbol{\Theta}\right) = \boldsymbol{h}\boldsymbol{\Theta}\boldsymbol{\upsilon}\text{,}
\end{equation}
for which $\boldsymbol{h} \in \mathbb{C}^{1 \times N}$ represents the channel between the \ac{UE} and \ac{BD-RIS} and $\boldsymbol{\upsilon} \in \mathbb{C}^N$ is the channel between \ac{BS} and \ac{BD-RIS}. Note that in practice the connections between \ac{BD-RIS} elements are usually implemented with components that have reciprocal impedance/admittance as, for example, resistors, capacitors, and inductors \cite[Sec.~III-A]{li:25}. Therefore, in addition to the unitary constraint ($\boldsymbol{\Theta}^\dagger \boldsymbol{\Theta} = \mathbf{I}$), hereon we also consider that the configuration matrix is symmetric, that is, $\boldsymbol{\Theta} = \boldsymbol{\Theta}^{\top}$. With that, the channel strength maximization problem for the overall channel $h_{\text{SISO}}\left(\boldsymbol{\Theta}\right)$ then becomes
\begin{equation}\label{eq:maxProbSISO}
    \begin{aligned}
        \max_{\boldsymbol{\Theta}} \ & \|\boldsymbol{h}\boldsymbol{\Theta}\boldsymbol{\upsilon}\|_2^2 \\
        \text{s.t. } \ & \boldsymbol{\Theta}^\dagger \boldsymbol{\Theta} = \mathbf{I}, \\
                       & \boldsymbol{\Theta} = \boldsymbol{\Theta}^{\top}.
    \end{aligned}
\end{equation}

By resorting to the Cauchy-Schwarz inequality, it becomes straightforward to demonstrate that the maximization of \eqref{eq:maxProbSISO} leads to the upper bound given by
\begin{subequations}\label{eq:upperbound}
    \begin{align}
        \|\boldsymbol{h}\boldsymbol{\Theta}\boldsymbol{\upsilon}\|_2^2 &\leq \|\boldsymbol{h}\|_2^2 \|\boldsymbol{\Theta}\boldsymbol{\upsilon}\|_2^2; \label{eq:upperbound_a} \\
        \max_{\boldsymbol{\Theta}} \|h_{\text{SISO}}\left(\boldsymbol{\Theta}\right)\|_2^2 &= \|\boldsymbol{h}\|_2^2 \|\boldsymbol{\upsilon}\|_2^2,\label{eq:upperbound_b}
    \end{align}
\end{subequations}
where we used the fact that $\boldsymbol{\Theta}^\dagger \boldsymbol{\Theta} = \mathbf{I}$. Observe that the equality holds in \eqref{eq:upperbound_a} if and only if the channel vectors are linearly dependent (same direction alignment). Consequently, we can also write that
\begin{equation}\label{eq:direction}
    \|\boldsymbol{\Theta}\boldsymbol{\bar{\upsilon}} - \boldsymbol{\bar{h}}^\dagger\|_2^2 = 2\left(1 - \Re\{\boldsymbol{\bar{h}}\boldsymbol{\Theta}\boldsymbol{\bar{\upsilon}}\}\right)\text{,}
\end{equation}
wherein $\boldsymbol{\bar{\upsilon}} = \boldsymbol{\upsilon}/\|\boldsymbol{\upsilon}\|_2$ and $\boldsymbol{\bar{h}} = \boldsymbol{h}/\|\boldsymbol{h}\|_2$, giving us a normalized parameter $0 \leq \Re\{\boldsymbol{\bar{h}} \boldsymbol{\Theta}\boldsymbol{\bar{\upsilon}}\} \leq 1$ that indicates the overall channel alignment, which is directly related to the channel strength. Therefore, we leverage this parameter in order to propose Algorithm \ref{alg:SISO}.
\begin{algorithm}[h!]
	\caption{Depth-First Tree Search with Pruning}\label{alg:SISO} 
        % \scriptsize
	\begin{algorithmic}[1]
            \Ensure $0 < \epsilon < 1$
            \Ensure  $m_L \gets z\tilde{\Theta}_{ij} /\|z\|_2\text{, } \tilde{\Theta}_{ij} \gets q \in_{\mathcal{U}}
            \mathcal{Q}$ \Comment{Root node}
            \Statex % blank line
            \Function{TreeSearch}{$l$, $m_l$, $\epsilon$, $\boldsymbol{h}$,
            $\boldsymbol{\upsilon}$, $\mathcal{Q}$}
                \Statex % blank line
                \Function{UniSym}{$\boldsymbol{\tilde{\Theta}}$} \label{line:uniSymFunc}
                    \State $\boldsymbol{\Theta}^\star \gets (\boldsymbol{\tilde{\Theta}} + \boldsymbol{\tilde{\Theta}}^\top)/2$
                    \State $(U\text{, } \Sigma\text{, } V) \gets \mathrm{SVD}
                        (\boldsymbol{\Theta}^\star)$
                    \State $\boldsymbol{\Theta}^\star \gets UV^\dagger$
                    \State \Return $\boldsymbol{\Theta}^\star$
                \EndFunction
                \Statex % blank line
                \If{$l = 0$}
                    \State \Return $\boldsymbol{\tilde{\Theta}}$
                \EndIf
                \Statex % blank line
                \State $\beta \gets 0$
                \State $l \mapsto \left(i,j\right)$\label{line:levelMap}
                \For{$\tilde{\Theta}_{ij} \in \mathcal{Q}$}\label{line:searchInit} \Comment{Explore branch}
                    \If{$i = j$}
                        \State $z \gets h_i\upsilon_j$ \label{line:diagInit}
                        \State $m_l \gets m_{l+1} + z\tilde{\Theta}_{ii}/\|z\|_2$
                        \State $r \gets \|m_l/2\|_2^2$ \label{line:diagEnd}
                    \Else
                        \State $\{\tilde{\Theta}_{ji}^p\}_{p=1}^\varrho \gets q_p$ \label{line:offdiagInit}
                        \State $\boldsymbol{\Theta}^p \gets \Call{UniSym}{\boldsymbol{\tilde{\Theta}}^p}, \ \forall p$ \label{line:uniSym}
                        \State $m^p_l \gets \boldsymbol{h}\boldsymbol{\Theta}^p\boldsymbol{\upsilon}/\|\boldsymbol{h}\|_2 \|\boldsymbol{\upsilon}\|_2 ,  \ \forall p$ \label{line:calc}
                        \State $r \gets \max{\ \Re\{m^1_l,m^2_l,\dots,m^\varrho_l\}}$ \label{line:offdiagMax}
                        \State $q_{\text{max}} \gets \underset{\forall \mathcal{Q}}{\arg \max}{\ \Re\{m^1_l,m^2_l,\dots,m^\varrho_l\}}$ \label{line:offdiagEnd}
                    \EndIf
                    \If{$r > \epsilon$} \label{line:prunInit} \Comment{Leaf node}
                        \State $\tilde{\Theta}_{ij} \gets q$
                        \State $\tilde{\Theta}_{ji} \gets q_{\text{max}}$
                        \State \Call{TreeSearch}{$l - 1$, $m_l$, $\epsilon$, $\boldsymbol{h}$, $\boldsymbol{\upsilon}$, $\mathcal{Q}$} \label{line:prunEnd}
                    \ElsIf{$r > \beta$} \label{line:exhInit}
                        \State $\tilde{\Theta}_{ij} \gets q$
                        \State $\tilde{\Theta}_{ji} \gets q_{\text{max}}$
                        \State $\beta \gets r$ \label{line:exhEnd}
                    \EndIf
                \EndFor \label{line:searchEnd}
            \EndFunction
	\end{algorithmic} 
\end{algorithm}

In Algorithm \ref{alg:SISO}, a tree search is performed over the discrete set of phase-shifts for each entry $\Theta_{ij}$ of the \ac{BD-RIS} configuration matrix. Throughout this section we consider the so-called fully-connected \ac{BD-RIS} matrix, for which each element is interconnected to all remaining $N - 1$ elements. Furthermore, the discrete set of phase-shifts is denoted by $\mathcal{Q} \in [-\pi,\pi)$, where $\pi/2^{(\log_2 \lvert\mathcal{Q}\rvert - 1)}$ is the step size. Note that the computational complexity of an exhaustive search over all possible phase-shift combinations is $\mathcal{O}(\lvert\mathcal{Q}\rvert^{N^2})$ in general, which is prohibitively complex. The tree search of Algorithm \ref{alg:SISO}, however, lowers this complexity significantly, since the depth-first search approach is employed and also pruning is used. More details about the computational complexity are given in Section~\ref{sec:numResult}.

More specifically, observe in Algorithm \ref{alg:SISO} from lines (\ref{line:searchInit})--(\ref{line:searchEnd}), that for each entry of the discretized \ac{BD-RIS} configuration matrix, $\tilde{\Theta}_{ij}$, a candidate phase-shift $q$ from the set $\mathcal{Q}$ is evaluated. The variable $0 \leq r \leq 1$ represents its suitability in the sense of maximizing the projection of $\boldsymbol{\bar{\upsilon}}$ in relation to $\boldsymbol{\bar{h}}$ and, consequently, of maximizing the objective function in \eqref{eq:maxProbSISO}. This process can be seen as the $l$-th, $l \in \left[0\text{,}1\text{,}\dots\text{,}L\right]$, tree branch exploration stage, so that when $r > \epsilon$ then the candidate is considered suitable (leaf node) and the process repeats itself for the next matrix entry or the next branch. Therefore, note that upon a suitable candidate, there occurs a jump to a deeper branch in the tree\footnote{A total of $L = N(N + 1)/2$ levels are explored.}, hence the depth-first search approach. Moreover, see that the lines (\ref{line:prunInit})--(\ref{line:prunEnd}) describe the pruning procedure, whereby other potential phase-shift candidates are discarded if the current candidate is suitable enough. This is determined by the value stored in $\epsilon$, which is typically assigned heuristically. If no suitable candidate is found, then an exhaustive search of the branch takes place and the best candidate (largest $r$) is chosen \commenttxt{(see lines (\ref{line:exhInit})--(\ref{line:exhEnd}))}.

The outline of Algorithm \ref{alg:SISO} given above can be generally applied to any conventional depth-first tree search algorithm with pruning. In the following, more specialized details are clarified regarding the particulars of the proposed algorithm and how they relate to problem \eqref{eq:maxProbSISO}.

\subsection{\ac{BD-RIS} Configuration: Proposed Algorithm}

Firstly, Algorithm \ref{alg:SISO} must be initialized, and this is ensured by the attribution of a randomly (uniform  distribution) chosen phase-shift from the set $\mathcal{Q}$ to a given entry of the \ac{BD-RIS} configuration matrix $\boldsymbol{\tilde{\Theta}}$. This operation defines the so-called root node on the first level of the tree. Moreover, each tree level index, $l$, is being mapped to a given row and column of $\boldsymbol{\tilde{\Theta}}$, respectively (see line (\ref{line:levelMap})), so that the operations of lines (\ref{line:diagInit})--(\ref{line:diagEnd}) are executed for diagonal entries and the ones from lines (\ref{line:offdiagInit}) through (\ref{line:offdiagEnd}) for off-diagonal entries. In principle, any arbitrary mapping between tree level and matrix entries could be chosen, but here we operate with diagonal entries first. Therefore, observe that for diagonal entries the operations are performed for each entry separately, where each discrete phase-shift is weighted by the respective normalized channel gains, as described in \eqref{eq:direction}. These operations would suffice for a diagonal \ac{RIS} configuration matrix, making the maximization problem of \eqref{eq:maxProbSISO} reasonably trivial in this case. However, new implications arise whilst configuring the off-diagonal entries of a \ac{BD-RIS} matrix.

\begin{figure}[h!]
	\centering
	\includegraphics[width=0.75\columnwidth,keepaspectratio]{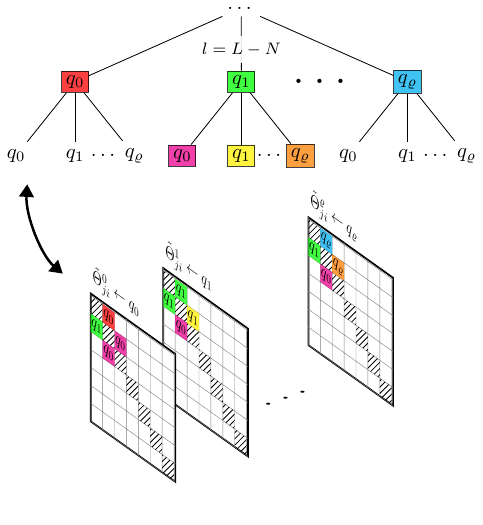}
	\caption{An example of the tree search exploration executed by Algorithm \ref{alg:SISO}.}\label{fig:lvlmapPlot}
\end{figure}

For off-diagonal entries, the first step is the attribution of all phase-shifts from $\mathcal{Q}$ to the reciprocal entries, $\tilde{\Theta}_{ji}$. Observe that line (\ref{line:offdiagInit}) of Algorithm \ref{alg:SISO} describes this operation. This means that all remaining operations are computed considering multiple matrices, the difference between them being only the value of the reciprocal entry. Consequently, on lines (\ref{line:offdiagMax}) and (\ref{line:offdiagEnd}) the maximum value among the multiple values stored in $r$ is computed, and the phase-shift in $\mathcal{Q}$ for which this maximum occurs, respectively. To elaborate, see in Figure~\ref{fig:lvlmapPlot} that the general tree search exploration executed by Algorithm \ref{alg:SISO} is being visually represented. For the sake of the example, consider that the candidate phase-shift for the first off-diagonal entry has already been chosen (diagonal entries also, necessarily), that is, say $q_1$ was chosen, where $\boldsymbol{q} \in \left[q_0\text{,}q_1\text{,}\dots\text{,}q_\varrho \right]$; $\varrho = \lvert\mathcal{Q}\rvert$. Therefore, as represented in Figure~\ref{fig:lvlmapPlot}, multiple matrices are evaluated and the one yielding the largest value for $r$ is chosen, as explained before. Moreover, observe in Figure~\ref{fig:lvlmapPlot} that each level or branch of the tree is mapped into an entry of the matrix, where the entries pertaining to the diagonals are explored in sequence.

In order to compute $r$ and evaluate the candidate phase-shift suitability, it also becomes necessary to observe the constraints of \eqref{eq:maxProbSISO}. Therefore, at line (\ref{line:uniSym}) a function is called to calculate the closest symmetric unitary projection \cite{fang:24} of the discrete matrix $\boldsymbol{\tilde{\Theta}}$. This is done by first realizing that the closest symmetric projection for a square matrix can be defined as \cite{golub:13,horn:12}
\begin{equation}\label{eq:minSym}
    \min_{\boldsymbol{\Theta} = \boldsymbol{\Theta}^{\top}} \ \|\boldsymbol{\tilde{\Theta}} - \boldsymbol{\Theta}\|_F^2 \text{,}
\end{equation}
for which $\boldsymbol{\Theta}^\star = (\boldsymbol{\tilde{\Theta}} + \boldsymbol{\tilde{\Theta}}^\top)/2$ minimizes \eqref{eq:minSym}. Likewise, the closest unitary projection for a square matrix, also widely known as the orthogonal Procrustes problem \cite{golub:13} (see Sec. 6.4.1), is given by
\begin{equation}\label{eq:minUni}
    \min_{\boldsymbol{\Theta}^\dagger \boldsymbol{\Theta} = \mathbf{I}} \ \|\boldsymbol{\tilde{\Theta}} - \boldsymbol{A}\boldsymbol{\Theta}\|_F^2 \text{,}
\end{equation}
where here $\boldsymbol{A} = \boldsymbol{I}$ and the matrix that minimizes \eqref{eq:minUni} is obtained via the \ac{SVD} of $\boldsymbol{A}^\top \boldsymbol{\tilde{\Theta}}$. Finally, after the function (see line (\ref{line:uniSymFunc})) returns the unitary symmetric matrix, then the variable $r$ is computed as described in \eqref{eq:direction}.

\subsubsection{Proposed Algorithm Validation}
Computational simulations are employed to validate Algorithm \ref{alg:SISO}, with the numerical results obtained from these simulations being briefly discussed in this subsection. We consider that the complex channel gains are drawn from a complex normal Gaussian distribution, such that $h_n,\upsilon_n \sim \mathcal{N}(0, 1/\sqrt{2})$, $\forall n$. \commenttxt{Note that the \ac{CSI} is assumed to be ideal and that all \ac{BD-RIS} elements are configured within each channel realization \cite[Sec.~II]{li:25}. All simulation parameters used to generate the numerical results are presented in Table~\ref{tbl:SISOsimParam}.}
\begin{table}[h!]
    \caption{Simulation parameters considering the \ac{SISO} system.}\label{tbl:SISOsimParam}
    \centering
    % \normalsize
    \setlength{\tabcolsep}{1mm}
    % \resizebox{0.66\linewidth}{!}{%
    \begin{tabular}{m{0.45\columnwidth} m{0.45\columnwidth}} \toprule
        \textbf{Parameter} & \textbf{Value} \\
        \cmidrule(lr){1-1}\cmidrule(lr){2-2}
        Channel Realizations & $500$ \\[0.5em]
        Channel Model & Complex Normal Gaussian \\[0.5em]
        \ac{BD-RIS} Upper Bound & $N^2$ \\[0.5em]
        Diagonal \ac{RIS} Upper Bound & $N + N(N - 1)\pi^2/16$ \\[0.5em]
        $N$ & $[4\text{, }9\text{, }16\text{, }25\text{, }36\text{, }49\text{, }64]$ \\[0.5em]
        $\epsilon$ & $[0.1\text{, }0.5]$ \\[0.5em] 
        $\log_2\lvert\mathcal{Q}\rvert$ & $[1\text{, }4]$ \\ \bottomrule
    \end{tabular}
    % }
\end{table}

\begin{figure}[h!]
	\centering
	\includegraphics[width=0.66\columnwidth,keepaspectratio]{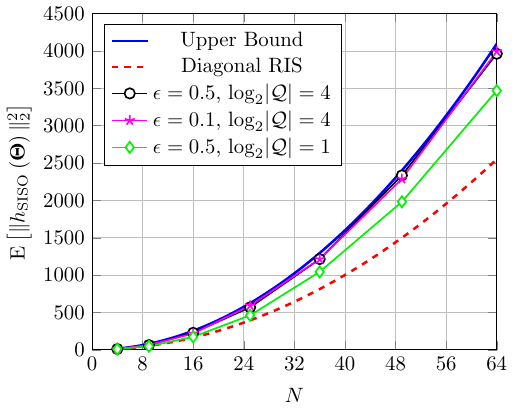}
	\caption{Average channel gain achieved by the proposed \ac{BD-RIS} configuration method of Algorithm \ref{alg:SISO}.}\label{fig:SISOPlot}
\end{figure}

Therefore, Figure \ref{fig:SISOPlot}
illustrates the average channel gain achieved by the proposed \ac{BD-RIS} configuration method of Algorithm \ref{alg:SISO}, where different parametrization for $\epsilon$ and the set size $\lvert\mathcal{Q}\rvert$ are considered. These results are plotted against different \ac{BD-RIS} matrices sizes and they are also validated in comparison to the analytical upper bound \cite[Sec.~II]{li:25} of both the \ac{BD-RIS} and the diagonal \ac{RIS}.

It can be concluded from Figure \ref{fig:SISOPlot} that running Algorithm \ref{alg:SISO} with a set size of $\log_2\lvert\mathcal{Q}\rvert = 4$ is already sufficient for obtaining results that practically achieve the upper bound, regardless of the configuration for $\epsilon$. See also in Figure \ref{fig:SISOPlot}, that for a smaller set size of $\log_2\lvert\mathcal{Q}\rvert = 1$ the average gain is noticeably lower. This shows that the \ac{BD-RIS} configuration with Algorithm \ref{alg:SISO} allows for flexible trade-offs between computational complexity and performance. Nevertheless, with the above discussion it becomes clear that the performance of Algorithm \ref{alg:SISO} is now validated, albeit for the specialized \ac{SISO} system. In the next section the \ac{BD-RIS} configuration for the \ac{MU-MISO} system is discussed and a more general algorithm is proposed, for which Algorithm \ref{alg:SISO} serves as a foundation.

\section{\ac{BD-RIS} Configuration: \ac{MU-MISO} System}\label{sec:MUMISO}
With the overall channel for the \ac{MU-MISO} system being given by \eqref{eq:chModelSISO}, we likewise have the channel strength maximization problem being now represented by
\begin{equation}\label{eq:maxProbMUMISO}
    \begin{aligned}
        \max_{\boldsymbol{\Theta}} \ & \|\boldsymbol{G}^\dagger + \boldsymbol{H}^\dagger \boldsymbol{\Theta}\boldsymbol{\Upsilon}\|_F^2 \\
        \text{s.t. } \ & \boldsymbol{\Theta}^\dagger \boldsymbol{\Theta} = \mathbf{I}, \\
                       & \boldsymbol{\Theta} = \boldsymbol{\Theta}^{\top}.
    \end{aligned}
\end{equation}
Having a closer look at the objective function of \eqref{eq:maxProbMUMISO} reveals that
\begin{equation}\label{eq:objFunc}
    \|\boldsymbol{G}^\dagger + \boldsymbol{H}^\dagger \boldsymbol{\Theta}\boldsymbol{\Upsilon}\|_F^2 = \|\boldsymbol{G}\|_F^2 + 2\Re\{\trace{(\boldsymbol{\Theta}\boldsymbol{X})}\} +  \trace{(\boldsymbol{\Theta}^\dagger \boldsymbol{Y}\boldsymbol{\Theta}\boldsymbol{Z})},
\end{equation}
where $\boldsymbol{X} = \boldsymbol{\Upsilon}\boldsymbol{G}\boldsymbol{H}^\dagger$, $\boldsymbol{Y} = \boldsymbol{H}\boldsymbol{H}^\dagger$ and $\boldsymbol{Z} = \boldsymbol{\Upsilon}\boldsymbol{\Upsilon}^\dagger$. Therefore, similarly to Section \ref{sec:SISO}, we leverage \eqref{eq:objFunc} and propose Algorithm \ref{alg:MUMISO}.  
\begin{algorithm}[h!]
	\caption{\ac{BD-RIS} Configuration for \ac{MU-MISO} Systems}\label{alg:MUMISO} 
        % \scriptsize
	\begin{algorithmic}[1]
            \Ensure $0 < \epsilon < 1$, $\rho > 0$, $d > 0$
            \Ensure $r^\prime \gets 0$, $c \gets 0$
            \Ensure $\boldsymbol{X} \gets \boldsymbol{\Upsilon}\boldsymbol{G}\boldsymbol{H}^\dagger$, $\boldsymbol{Y} \gets \boldsymbol{H}\boldsymbol{H}^\dagger$, $\boldsymbol{Z} \gets \boldsymbol{\Upsilon}\boldsymbol{\Upsilon}^\dagger$
            % \Statex % blank line
            \Function{TreeSearch}{$l$, $\epsilon$, $\rho$, $d$, $r^\prime$, $c$, $\boldsymbol{X}$, $\boldsymbol{Y}$, $\boldsymbol{Z}$, $\mathcal{Q}$}
            % \Statex % blank line
                \Function{VarCalc}{$\boldsymbol{X}$, $\boldsymbol{Y}$, $\boldsymbol{Z}$, $\boldsymbol{\Phi}$} \label{line:varCalc}
                    \If{$\|\boldsymbol{G}\|_F^2 = 0$} \Comment{Obstructed channel}
                            \State $z \gets \|\boldsymbol{Y}\|_F\|\boldsymbol{Z}\|_F$
                            \State $m_l \gets \trace{(\boldsymbol{\Phi}^\dagger\boldsymbol{Y}\boldsymbol{\Phi}\boldsymbol{Z})}$ \label{line:mOb}
                            \State $w \gets (\Re\{m_l\} + z)(2z)^{-1}$ \label{line:wOb}
                        \Else
                            \State $m_l \gets \trace{(\boldsymbol{\Phi}\boldsymbol{X})}$
                            \State $w \gets (\Re\{m_l\} + \|\boldsymbol{X}\|_*)(2\|\boldsymbol{X}\|_*)^{-1}$ \label{line:wUn}
                        \EndIf
                    \State \Return $w$
                \EndFunction
                % \Statex % blank line
                \Function{UniSym}{$\boldsymbol{\tilde{\Theta}}$} \label{line:unisym}
                    \State $\boldsymbol{\Theta}^\star \gets (\boldsymbol{\tilde{\Theta}} + \boldsymbol{\tilde{\Theta}}^\top)/2$
                    \State $(U\text{, } \Sigma\text{, } V) \gets \mathrm{SVD}
                        (\boldsymbol{\Theta}^\star)$
                    \State $\boldsymbol{\Theta}^\star \gets UV^\dagger$
                    \State \Return $\boldsymbol{\Theta}^\star$
                \EndFunction
                \Statex % blank line
                \If{$l = 0$}
                    \State \Return $\boldsymbol{\tilde{\Theta}}$
                \EndIf
                \Statex % blank line
                \State $l \mapsto \left(i,j\right)$
                \For{$\tilde{\Theta}_{ij} \in \mathcal{Q}$}
                    \If{$i = j$}
                        \State $r \gets \Call{VarCalc}{\boldsymbol{X}, \boldsymbol{Y}, \boldsymbol{Z}, \boldsymbol{\tilde{\Theta}}}$
                    \Else
                        \State $\{\tilde{\Theta}_{ji}^p\}_{p=1}^\varrho \gets q_p$
                        \State $\boldsymbol{\Theta}^p \gets \Call{UniSym}{\boldsymbol{\tilde{\Theta}}^p} ,  \ \forall p$
                        \State $u^p \gets \Call{VarCalc}{\boldsymbol{X}, \boldsymbol{Y}, \boldsymbol{Z}, \boldsymbol{\Theta}^p} ,  \ \forall p$
                        \State $r \gets \max{\ u^1,u^2,\dots,u^\varrho}$
                        \State $q_{\text{max}} \gets \underset{\forall \mathcal{Q}}{\arg \max}{\ u^1,u^2,\dots,u^\varrho}$
                    \EndIf
                    \If{$r > \epsilon$}
                        \State $\tilde{\Theta}_{ij} \gets q$
                        \State $\tilde{\Theta}_{ji} \gets q_{\text{max}}$
                        \If{$\lvert r - r^\prime \rvert < \rho$} \label{line:branchPInit}
                            \State $c \gets c + 1$
                            \State $r^\prime \gets r$
                            \State \textbf{if} $c \geq d$ \textbf{then} $l \gets 1$
                            % \If{$c \geq d$}
                            %     \State $l \gets 1$
                            % \EndIf
                        \Else
                            \State $r^\prime \gets r$
                        \EndIf \label{line:branchPEnd}
                        \State \Call{TreeSearch}{$l - 1$, $\epsilon$, $\dots$, $\mathcal{Q}$}
                    \EndIf
                \EndFor
            \EndFunction
	\end{algorithmic} 
\end{algorithm}

\subsection{\ac{BD-RIS} Configuration: Proposed Algorithm}
Observe in Algorithm \ref{alg:MUMISO} that a dedicated function (see line (\ref{line:varCalc})) returns the variable, $w$, which indicates the candidate phase-shift suitability towards channel strength maximization. In this function, both the obstructed and unobstructed direct channel scenarios are taken into account. For the former scenario, the last term of \eqref{eq:objFunc} is used directly, since in this case $\|\boldsymbol{G}\|_F^2 = 0$, whereas for the latter the second term of \eqref{eq:objFunc} is employed. Moreover, in both of \commenttxtR{these} scenarios, the following upper bounds are considered for the variable normalization ($0 \leq w \leq 1$)
\begin{subequations}\label{eq:propRel}
    \begin{align}
       \lvert\trace{(\boldsymbol{\Theta}^\dagger \boldsymbol{Y}\boldsymbol{\Theta}\boldsymbol{Z})}\lvert &\leq \overset{\|\boldsymbol{Y}\|_F}{\overbrace{\|\boldsymbol{\Theta}^\dagger \boldsymbol{Y}\boldsymbol{\Theta}\|_F}} \|\boldsymbol{Z}\|_F; \label{eq:propRel_a} \\
        \Re\{\trace{(\boldsymbol{\Theta} \boldsymbol{X})}\} &\leq \|\boldsymbol{X}\|_*,\label{eq:propRel_b}
    \end{align}
\end{subequations}
wherein \eqref{eq:propRel_a} is the Cauchy-Schwarz inequality and \eqref{eq:propRel_b} is based on the von Neumann’s trace theorem \cite{horn:12} (see Sec. 7.4.1).

Algorithm \ref{alg:MUMISO} also avails of the so-called branch pruning, as shown in lines (\ref{line:branchPInit}) through (\ref{line:branchPEnd}). This means that the algorithm execution can be terminated prematurely if no significant improvement is obtained in that branch or level. The improvement is quantified by the performance difference between adjacent branches and the minimum improvement required is given by $\rho > 0$. \commenttxtR{Furthermore, the parameter $d$ allows for a delay in the premature or early termination of the algorithm execution, by simply counting how many times the performance improvement was not enough and subsequently terminating the algorithm execution if the counting extrapolates the value of $d$.} As a consequence, a more scalable cost in terms computational complexity can be achieved, since a range of $L \leq N(N + 1)/2$ levels are now allowed to be explored. 
% In practice, this entails a dynamically connected architecture for the \ac{BD-RIS}, with an admittance-switch network of $N(N - 1)/2$ switches that are able to dynamically activate or deactivate impedance/admittance components \cite[Sec.~III-D7]{li:25}. 

To conclude, the discussions provided in Section \ref{sec:SISO} should suffice for clarifications regarding the remaining parts of Algorithm \ref{alg:MUMISO}, given its similarity to Algorithm \ref{alg:SISO}.      

\section{Numerical Results}\label{sec:numResult}

\begin{figure}[h!]
	\centering
	\includegraphics[width=0.66\columnwidth,keepaspectratio]{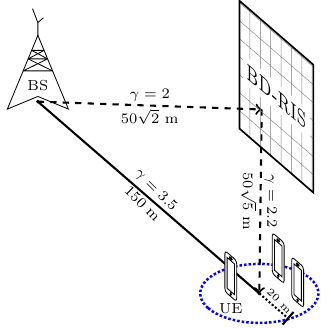}
	\caption{Depiction of the communications scenario investigated.}\label{fig:sysmodelPlot}
\end{figure}

In this work, we employ computational simulations to evaluate the performance of Algorithm~\ref{alg:MUMISO}. The communications scenario investigated is based on the approach of \cite{fang:24}, from which the performance of the proposed low-complexity \ac{BD-RIS} configuration (see Eq. (15) in \cite{fang:24}) is used as a baseline. \commenttxt{Additionally, we also provide results for the full-rank manifold optimization algorithm with phase optimization procedure\footnote{\commenttxt{To establish a benchmark, we opted for the best performing alternative of the manifold optimization procedures presented in \cite{santa:26}.}}, reported in \cite{santa:26}.} Therefore, all the channel coefficients are drawn from a normal complex Gaussian distribution and weighted by a distance-dependent path loss model given by: $\zeta (d) = \zeta_0 d^{-\gamma}$, where $\zeta_0 = -30$ dB is the reference path loss at a distance of $1$ meter, $d$ is the link distance, and $\gamma$ is the path loss exponent. The communications scenario depicted in Figure~\ref{fig:sysmodelPlot} 
shows the parameters used for the path loss exponent for each link of the overall channel and the respective link distances. Observe also in Figure~\ref{fig:sysmodelPlot}, that the $k$-th \ac{UE} is positioned randomly within a circle of $20$ meters radius ($150$ meters from the \ac{BS}), for which an uniform distribution of \acp{UE} is assumed. All simulation results were obtained by averaging $500$ channel realizations.

In the following, first the performance of Algorithm~\ref{alg:MUMISO} is evaluated followed by an analysis of its computational complexity. \commenttxt{All simulation parameters used to generate the numerical results are presented in Table~\ref{tbl:MUMISOsimParam}.
\begin{table}[h!]
    \caption{Simulation parameters for the \ac{MU-MISO} system.}\label{tbl:MUMISOsimParam}
    \centering
    % \normalsize
    \setlength{\tabcolsep}{1mm}
    % \resizebox{0.66\linewidth}{!}{%
    \begin{tabular}{m{0.45\columnwidth} m{0.45\columnwidth}} \toprule
        \textbf{Parameter} & \textbf{Value} \\
        \cmidrule(lr){1-1}\cmidrule(lr){2-2}
        Channel Realizations & $500$ \\[0.5em]
        Channel Model & Distance-dependent Gaussian \\[0.5em]
        Reference Path Loss & $\zeta_0 = -30$ dB ($1$ m) \\[0.5em]
        \ac{BS} $\rightarrow$ \ac{UE} & $d = 150$ m and $\gamma = 3.5$ \\[0.5em]
        \ac{BS} $\rightarrow$ \ac{BD-RIS} & $d = 50\sqrt{2}$ m and $\gamma = 2$ \\[0.5em]
        \ac{BD-RIS} $\rightarrow$ \ac{UE} & $d = 50\sqrt{5}$ m and $\gamma = 2.2$ \\[0.5em]
        $N$ & $[4\text{, }9\text{, }16\text{, }25\text{, }36\text{, }49\text{, }64]$ \\[0.5em]
        $L\times K$ & $4\times 4$, $4\times 8$, $8\times 8$ and $64\times 64$ \\[0.5em]
        $\epsilon$ & $[0.1\text{, }0.5\text{, }0.99]$ \\[0.5em] 
        $\log_2\lvert\mathcal{Q}\rvert$ & $4$ \\[0.5em]
        $\rho$ & $1\times10^{-4}$ \\[0.5em]
        Max. Iterations (Manifold) & $200$ \\[0.5em]
        Convergence Thresh. (Manifold) & $1\times10^{-4}$ \\\bottomrule
    \end{tabular}
    % }
\end{table}
\commenttxtR{Note that the channel parameters are defined identically to those in \cite{fang:24}, while the Manifold optimization parameters are the same as those used in \cite{santa:26}.}}

\subsection{Performance Results}
\begin{figure*}[h!]
	\centering
	\includegraphics[width=0.9\linewidth,keepaspectratio]{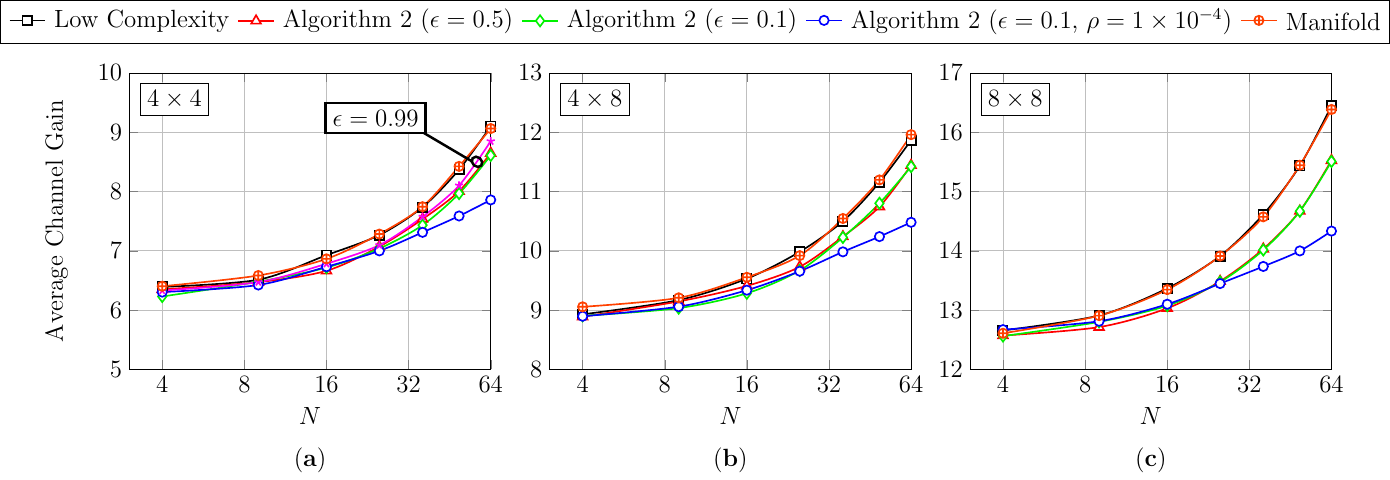}
	\caption{The average channel gain is computed for a range of different \ac{BD-RIS} matrix sizes, $N$, considering $4\times 4$ (a), $4\times 8$ (b) and $8\times 8$ (c) \ac{MU-MISO} systems. The results are generated for an unobstructed direct channel and Algorithm~\ref{alg:MUMISO} is configured with $\epsilon = [0.1\text{,}0.5]$ ($\epsilon = 0.99$ only for (a)), with branch pruning both enabled ($\rho = 1\times10^{-4}$) and disabled. The results from the low-complexity \ac{BD-RIS} configuration of \cite{fang:24} serve as a baseline\commenttxt{, whereas the manifold optimization reported in \cite{santa:26} serves as a benchmark.}}\label{fig:gainPlot}
\end{figure*}

\commenttxt{\subsubsection{Unobstructed Direct Channel}}
In Figure~\ref{fig:gainPlot}, the average channel gain is computed for a range of different \ac{BD-RIS} matrix sizes, $N$, and $L\times K$ \ac{MU-MISO} systems with $L\text{, }K = [4\text{, }8]$, where the scenario with the unobstructed direct channel is considered. Note also in Figure~\ref{fig:gainPlot} that results are shown for the Algorithm~\ref{alg:MUMISO} without the so-called branch pruning, for which values of $\epsilon = [0.1\text{, }0.5]$ are considered, and also otherwise set with branch pruning, where we have $\epsilon = 0.1$ and $\rho = 1\times10^{-4}$. The results for $\epsilon = 0.99$ (no branch pruning) are plotted only in Figure~\ref{fig:gainPlot} (a), since the computational complexity becomes prohibitive for this parameter setting. Moreover, recall that the proposed low-complexity \ac{BD-RIS} configuration of \cite{fang:24} is used as a baseline in this work. For this configuration, the \ac{BD-RIS} matrix is given by $\boldsymbol{\Theta} = \Call{UniSym}{\boldsymbol{H}\boldsymbol{G}^\dagger \boldsymbol{\Upsilon}^\dagger}$, where $\Call{UniSym}{\cdot}$ (line (\ref{line:unisym}) of Algorithm~\ref{alg:MUMISO}) returns the unitary symmetric matrix. \commenttxt{In addition, the results for the manifold optimization \cite{santa:26} are also presented in Figure~\ref{fig:gainPlot} and serve as a benchmark.} Finally, for all the remaining results discussed in this section the size of the discrete set of phase-shifts is fixed to $\log_2 \lvert\mathcal{Q}\rvert = 4$.

Firstly, it can be verified Figure~\ref{fig:gainPlot} (a) that the performance curve for the low-complexity method adheres well with the results reported in \cite{fang:24}, assuring a valid baseline for comparison. \commenttxt{Moreover, the results for the manifold optimization also adhere well to the baseline.} Consequently, observe in Figure~\ref{fig:gainPlot} that, in general, the performance of Algorithm~\ref{alg:MUMISO} is close to the results achieved by \commenttxt{both the baseline and benchmark}, with the exception being the performance penalty experienced by Algorithm~\ref{alg:MUMISO} with branch pruning. Most prominent for \ac{BD-RIS} matrix sizes of $N > 25$, this penalty is expected since in this case performance is being traded for computational complexity scalability. Note also in Figure~\ref{fig:gainPlot} (a) that the gain in performance is negligible when setting $\epsilon = 0.99$, that is, when Algorithm~\ref{alg:MUMISO} is practically forced to carry out an exhaustive branch search that is prohibitively complex.

\begin{figure*}[h!]
	\centering
	\includegraphics[width=0.9\linewidth,keepaspectratio]{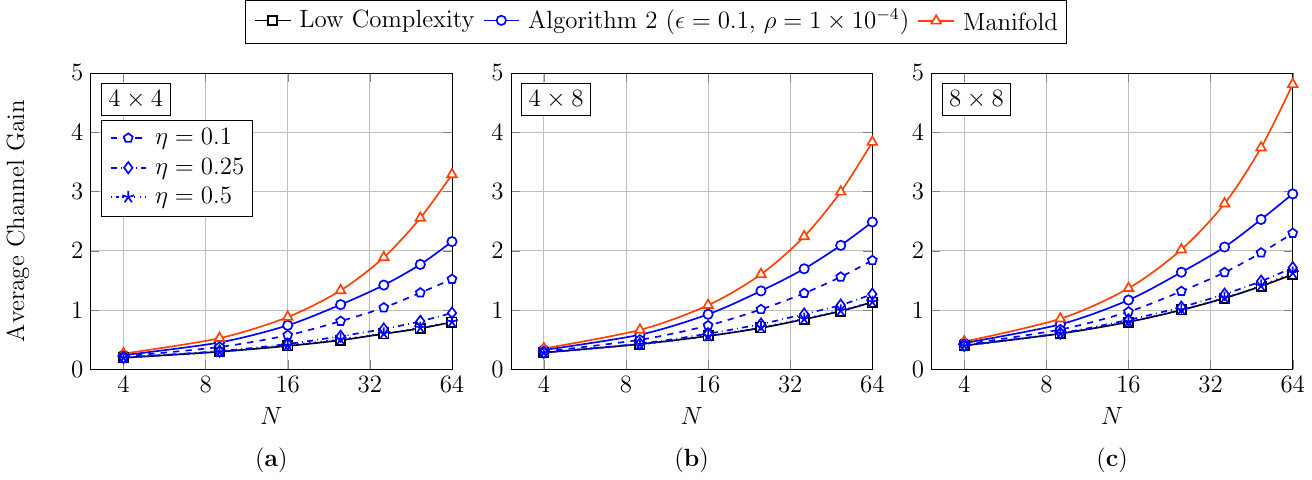}
	\caption{The average channel gain is computed for a range of different \ac{BD-RIS} matrix sizes, $N$, considering $4\times 4$ (a), $4\times 8$ (b) and $8\times 8$ (c) \ac{MU-MISO} systems. The results are generated for an obstructed direct channel and Algorithm~\ref{alg:MUMISO} is configured both with $\epsilon = 0.1$ and $\rho = 1\times10^{-4}$. The impact of imperfect \ac{CSI} on Algorithm~\ref{alg:MUMISO} is also evaluated, considering $\eta = 0.1$, $\eta = 0.25$ and $\eta = 0.5$.}\label{fig:obsgainPlot}
\end{figure*}

\commenttxt{\subsubsection{Obstructed Direct Channel}}
If instead the scenario with the obstructed direct channel ($\|\boldsymbol{G}\|_F^2 = 0$) is considered, then we have the results shown Figure~\ref{fig:obsgainPlot}.
For these results, the Algorithm~\ref{alg:MUMISO} with branch pruning is used, where all the remaining parameters are configured identically to Figure~\ref{fig:gainPlot}. \commenttxt{Furthermore, recall that all channel coefficients used by Algorithm~\ref{alg:MUMISO} are assumed to have been obtained through an ideal channel estimation process. Although this allows for a more straightforward discussion of the performance results, still the impact of non-ideal or imperfect \ac{CSI} should be discussed \cite{sun:25,sun:26}. Consequently, Figure~\ref{fig:obsgainPlot} also shows the results obtained when imperfect \ac{CSI} is taken into account for Algorithm~\ref{alg:MUMISO}. The estimation error is modeled as \cite{joudeh:14,pedro:24}
\begin{subequations}\label{eq:estErr}
    \begin{align}
        \hat{\mathbf{H}} &= \sqrt{1-\eta^2}\mathbf{H} + \eta\mathbf{E}_\mathbf{H}, \label{eq:estErr_a} \\ 
        \hat{\mathbf{\Upsilon}} &= \sqrt{1-\eta^2}\mathbf{\Upsilon} + \eta\mathbf{E}_{\mathbf{\Upsilon}}, \label{eq:estErr_b}
    \end{align}
\end{subequations}
where $\hat{\mathbf{H}}$ in \eqref{eq:estErr_a} represents the estimated complex channel gains between the \acp{UE} and all elements of the \ac{BD-RIS}. The same applies to the channel coefficients between \ac{BS} and \ac{BD-RIS} in \eqref{eq:estErr_b}. Moreover, we have the error matrices, $\mathbf{E}_\mathbf{H} \sim
\mathcal{CN}\!\left(\mathbf{0},\mathbf{I}\right)$ and $\mathbf{E}_{\mathbf{\Upsilon}} \sim
\mathcal{CN}\!\left(\mathbf{0},\mathbf{I}\right)$, which are assumed to be mutually independent and independent of the true channel matrices. Therefore, $\eta$ controls how degraded is the estimation in relation to the true channel coefficients, in which $\eta = 0$ represents the perfect \ac{CSI} and $\eta = 1$ models completely uncorrelated channel coefficients.   
} 

Observe in Figure~\ref{fig:obsgainPlot} that the performance of Algorithm~\ref{alg:MUMISO} \commenttxt{(perfect \ac{CSI})} is significantly better than the baseline low-complexity method. The poor performance of the baseline is a consequence of the first-order approximation used in \cite{fang:24} to obtain the low-complex configuration of the \ac{BD-RIS}. \commenttxt{On the other hand, the Algorithm~\ref{alg:MUMISO} performance is slightly worse than that obtained by the manifold optimization.} Therefore, Algorithm~\ref{alg:MUMISO} is able to maintain \commenttxt{comparable} levels of performance \commenttxt{in relation to the benchmark, whilst showing an improvement in relation to the baseline} across different communications scenarios. \commenttxt{From Figure~\ref{fig:obsgainPlot} one can also conclude that the imperfect \ac{CSI} has a sensible impact on the performance of Algorithm~\ref{alg:MUMISO}, particularly for $\eta \geq 0.25$. Although the estimation error model of \eqref{eq:estErr} can be considered too pessimistic, these results show nevertheless that channel estimation may play an important role for such \ac{BD-RIS} configuration methods.} In the next subsection we demonstrate that Algorithm~\ref{alg:MUMISO} can also be very scalable terms of computational complexity.

\subsection{\commenttxt{Convergence and} Computational Complexity Results}

\begin{figure}[h!]
	\centering
	\includegraphics[width=0.66\columnwidth,keepaspectratio]{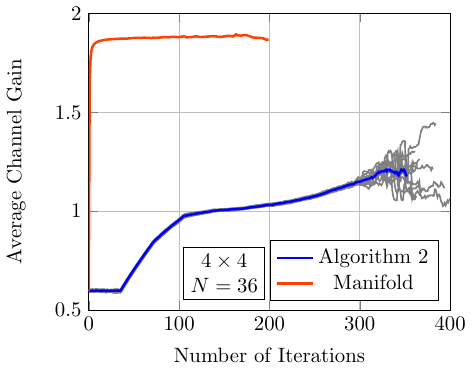}
	\caption{The convergence behavior for Algorithm~\ref{alg:MUMISO} and the manifold optimization.}\label{fig:IterPlot}
\end{figure}

\commenttxt{\subsubsection{Convergence Results}}
\commenttxt{The values for the average channel gain for each iteration of Algorithm~\ref{alg:MUMISO} and the manifold optimization of \cite{santa:26} are illustrated in Figure~\ref{fig:IterPlot}, unveiling the average convergence behavior of these \ac{BD-RIS} configuration algorithms. We consider a representative scenario with a \ac{BD-RIS} matrix size of $N = 36$ and an $4\times 4$ \ac{MU-MISO} system, given that no relevant changes in the convergence behavior were observed for other scenarios. Moreover, we assume that the direct channel is obstructed, which in context of this work represents the case where the baseline low-complexity method presents a poor performance. Otherwise, the baseline achieves the reference performance with a single iteration of the first-order approximation. Therefore, note that here we employ the same parameter configuration for Algorithm~\ref{alg:MUMISO} as used in Figure~\ref{fig:obsgainPlot}. \commenttxtR{Finally, note also that the average convergence behavior of Algorithm~\ref{alg:MUMISO} is drawn from multiple convergence curves, which are being illustrated (grayed out) in Figure~\ref{fig:IterPlot}. In order to obtain a statistically consistent average convergence behavior, we capped the maximum number of iterations at $352$, since we observed that only in rare occurrences the number of iterations were much greater than this.}   

Through Figure~\ref{fig:IterPlot} it can be verified that both algorithms have an increasing average channel gain with more iterations, showing a clear trend of improvement with each iteration. However, Algorithm~\ref{alg:MUMISO} shows a slower convergence behavior in comparison with the manifold optimization. This is counterbalanced by Algorithm~\ref{alg:MUMISO}, since it exhibits a lower computational complexity for each iteration, as will be discussed in the next subsection.}

\commenttxt{\subsubsection{Complexity Results}}
The asymptotic computational complexity of Algorithm~\ref{alg:MUMISO} is $\mathcal{O}(\lvert\mathcal{Q}\rvert N^5)$ when considering an exhaustive search for each branch ($\epsilon = 1$), while also assuming that branch pruning is disabled ($\rho = 0$). This can be seen as an upper bound on the complexity of Algorithm~\ref{alg:MUMISO}. Note that $\Call{VarCalc}{\cdot}$ (line (\ref{line:varCalc})) and $\Call{UniSym}{\cdot}$ (line (\ref{line:unisym})) cost $\mathcal{O}(N^3)$ each \cite{fang:24} and that the for loop runs a maximum of $\lvert\mathcal{Q}\rvert$ iterations. Consequently, by also knowing that Algorithm~\ref{alg:MUMISO} is called recursively $N(N + 1)/2$ times when branch pruning is not set, then the total computational complexity can be found to be approximately $\mathcal{O}(\lvert\mathcal{Q}\rvert N^4 (N + 1)/2) \sim \mathcal{O}(\lvert\mathcal{Q}\rvert N^5)$. \commenttxt{Furthermore, note also that the computational complexity per iteration of Algorithm~\ref{alg:MUMISO} is lower in comparison to the manifold optimization, which exhibits a computational complexity of $\mathcal{O}(NK^2 +N^2K +N^3)$ (with $L = K$) per iteration \cite{santa:26}. We assume that the multiple matrices computations for each reciprocal entry can be done in parallel, thus not increasing the cost per iteration of $\mathcal{O}(N^3)$ for Algorithm~\ref{alg:MUMISO}.}  

\begin{figure*}[h!]
	\centering
	\includegraphics[width=0.9\linewidth,keepaspectratio]{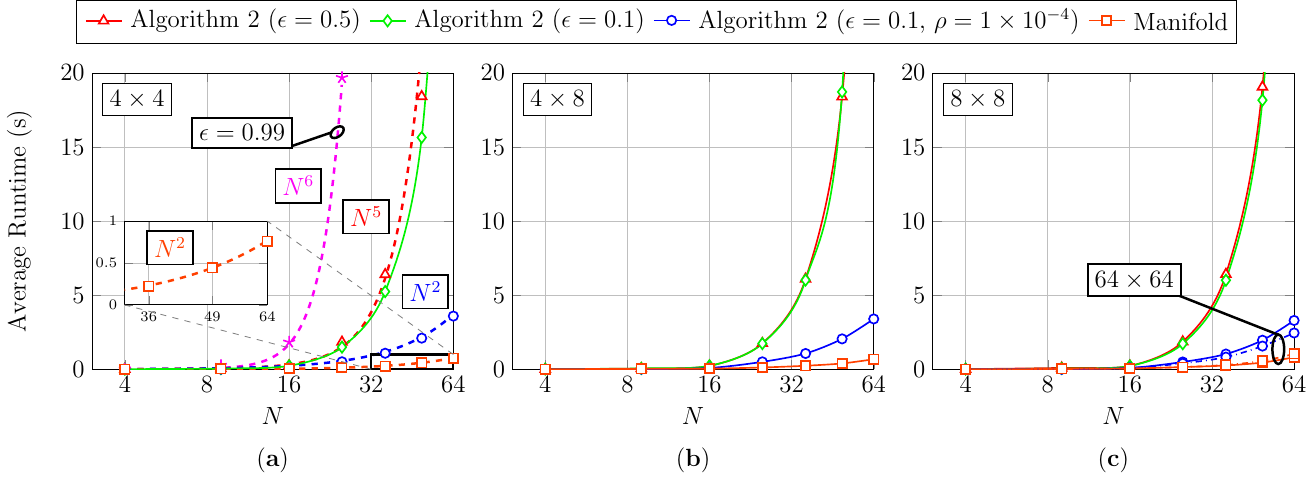}
	\caption{Computational complexity of Algorithm~\ref{alg:MUMISO} \commenttxt{and for the manifold optimization} based on the average runtime. The settings used for Algorithm~\ref{alg:MUMISO} are identical to those reported for Figure~\ref{fig:gainPlot}.}\label{fig:runtimePlot}
\end{figure*}

In order to provide a numerical validation of the computational complexity, the average runtime of Algorithm~\ref{alg:MUMISO} \commenttxt{and the manifold optimization are} computed via computational simulations. We employed a consumer computer with the \commenttxt{AMD Ryzen(TM) $5$ $3600$XT} processor and \commenttxt{$32$} GB of RAM, running the \commenttxt{Ubuntu $24.04.4$ LTS} operating system. Figure~\ref{fig:runtimePlot}
shows the average runtime in seconds (s) for different \ac{BD-RIS} matrix sizes and \ac{MU-MISO} system configurations, where the settings used for Algorithm~\ref{alg:MUMISO} are the same as in Figure~\ref{fig:gainPlot}. Moreover, see that the dashed lines in Figure~\ref{fig:runtimePlot} (a) are the result of an approximate polynomial fitting of the numerical results obtained for the runtime. Only the polynomial terms with the largest magnitude are highlighted, particularly because the asymptotic trend of the computational complexity is mainly of interest here.  

Therefore, it can be concluded from Figure~\ref{fig:runtimePlot} (a) that the numerical results adhere well to the asymptotic computational complexity, especially for $\epsilon = 0.99$, in which $\mathcal{O}(\lvert\mathcal{Q}\rvert N^5)$ represents a tighter bound. In other words, recall that $\lvert\mathcal{Q}\rvert = 16$ is being used, meaning that for $N \sim 16$ we actually have a cost of approximately $\mathcal{O}(N^6)$. As the pruning parameters are set with appropriate values the computational complexity can diminish considerably, as observed for the Algorithm~\ref{alg:MUMISO} with branch pruning, which presents a cost of approximately $\mathcal{O}(N^2)$. \commenttxt{More importantly, note that this cost is comparable to that exhibited by the manifold optimization. Although the asymptotic computational complexity per iteration of the manifold optimization is dominated by a cubic term \cite{santa:26} (as for Algorithm~\ref{alg:MUMISO}), keep in mind that these numerical results are obtained through efficient implementations of numerical linear algebra libraries. Therefore, it is expected that the runtime results could present lower values for computational complexity, as numerical structures of matrices are taken advantage of for faster computations.} Furthermore, \commenttxt{observe that} in general the computational complexity is practically invariant for different \ac{MU-MISO} systems\commenttxt{, particularly when considering the results shown in Figure~\ref{fig:runtimePlot} (c) for a $64\times 64$ \ac{MU-MISO} system}.
% Important to note that the low-complexity \ac{BD-RIS} configuration of \cite{fang:24} is omitted in Figure~\ref{fig:runtimePlot} since its runtime values are too small to be considered.

\begin{figure*}[hbtp]
	\centering
	\includegraphics[width=0.99\linewidth,keepaspectratio]{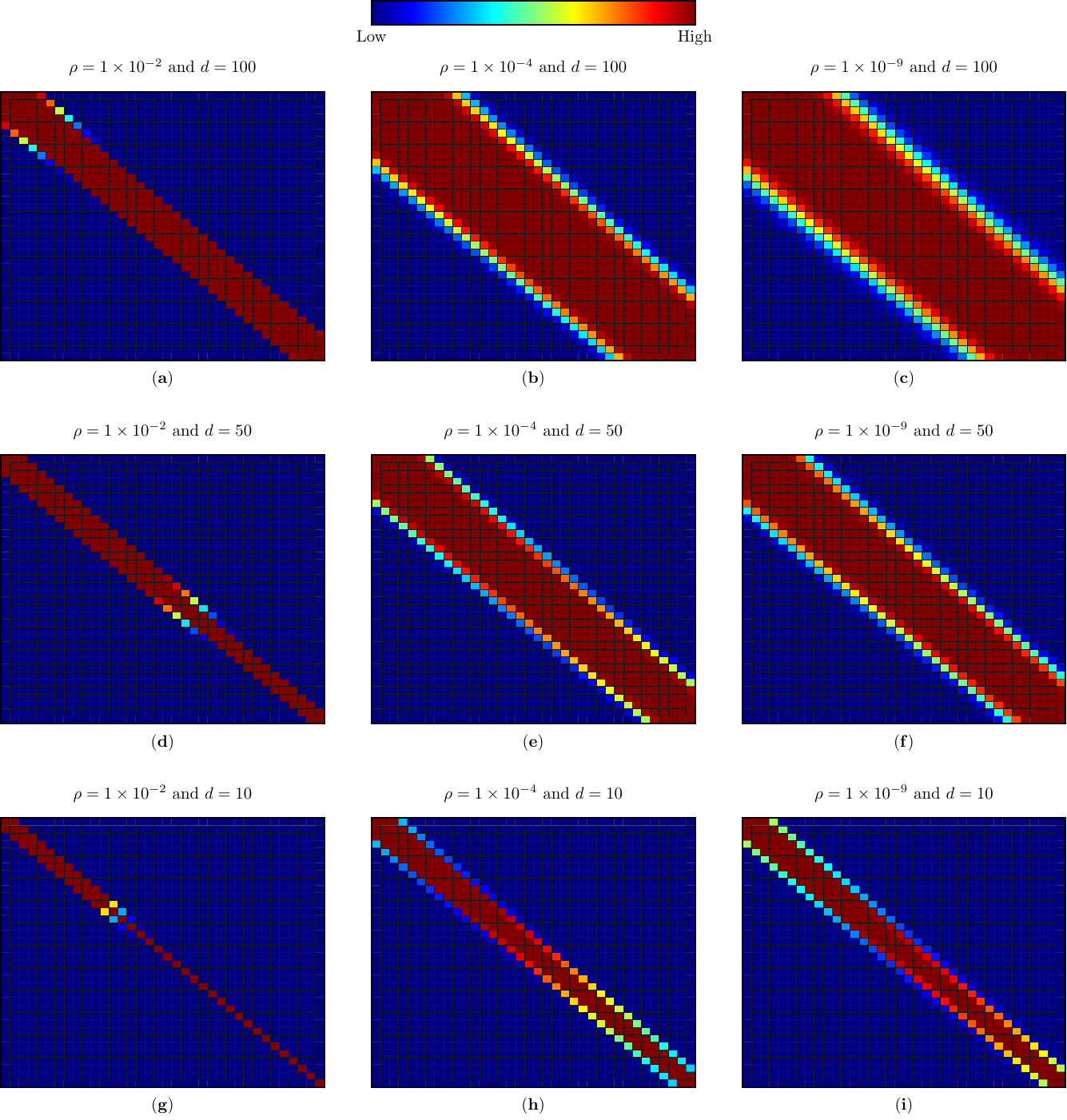}
	\caption{The impact of the branch pruning parameters, $\rho$ and $d$, on the average number of levels explored in Algorithm~\ref{alg:MUMISO}. The average usage of the discrete $36\times 36$ matrix entries is indicated by colors: blue represents low usage and red high usage (this figure is better visualized in colors).}\label{fig:heatmapPlot}
\end{figure*}

The impact of the branch pruning parameter, $\rho$, becomes more evident in Figure~\ref{fig:heatmapPlot}, where values for \mbox{$\rho = [1\times10^{-2}\text{, }1\times10^{-4}\text{, }1\times10^{-9}]$} and \mbox{$d = [10\text{, }50\text{, }100]$} are evaluated. Moreover, notice in Figure~\ref{fig:heatmapPlot} that the discretized \ac{BD-RIS} configuration matrix, $\boldsymbol{\tilde{\Theta}}  \in \mathcal{Q}^{36 \times 36}$, is represented by colors that indicate the average usage of its entries. Recall that Algorithm~\ref{alg:MUMISO} employs a mapping between tree level and matrix entries and, throughout this work as well as in Figure~\ref{fig:heatmapPlot}, we consider a diagonal mapping, where diagonal entries are mapped successively as more levels are explored. Therefore, observe in Figure~\ref{fig:heatmapPlot} (c) that more entries are used on average for smaller values of $\rho$ and larger values for $d$, with the opposite being demonstrated in Figure~\ref{fig:heatmapPlot} (g). In general, Figure~\ref{fig:heatmapPlot} shows that the number of levels (matrix entries) explored by Algorithm~\ref{alg:MUMISO} can be fine-tuned with great flexibility, consequently allowing for a satisfactory trade-off between performance and computational complexity across different communications scenarios.

\commenttxtR{In conclusion, although the average runtime of Algorithm~\ref{alg:MUMISO} shown in Figure~\ref{fig:runtimePlot} is higher in comparison to the manifold optimization (from $\sim N > 25$ onward), the difference is on the order of a few seconds. Furthermore, recall that the asymptotic complexity varies with the square of the \ac{BD-RIS} matrix size for both configuration methods. Therefore, the performance as well as the computational complexity results can be regarded as roughly similar or comparable between the configuration methods investigated.}

\section{Future Research and Open Problems}\label{sec:future}
We have so far demonstrated that Algorithm~\ref{alg:MUMISO} is an efficient, yet scalable alternative for \ac{BD-RIS} configuration in \ac{MU-MISO} communication systems. However, there are still several aspects of Algorithm~\ref{alg:MUMISO} that remain to be explored and improved. Since the main objective of this work was to lay the groundwork for \ac{BD-RIS} configuration with tree search algorithms, we expect that other research problems could be further developed in future works. Therefore, in the following, we provide a brief discussion of improvements to different aspects and functionalities of Algorithm~\ref{alg:MUMISO} that may be of interest.        

\subsection{Active Beamforming Integration}
Typically, the \ac{BD-RIS} configuration is resolved in conjunction with the active beamforming configuration, given that, prior to the signal transmission, active beamforming vectors can be employed at the \ac{BS} in order to improve the \ac{SINR} for all \acp{UE}. As a consequence, the channel strength maximization problem must be now solved jointly with the sum-rate maximization problem. Therefore, solutions for integrating Algorithm~\ref{alg:MUMISO} with active beamforming are needed where the scalability and efficiency properties of this algorithm are maintained.  

% \subsection{Imperfect \ac{CSI}}
% All channel coefficients used by Algorithm~\ref{alg:MUMISO} are assumed to have been obtained through an ideal channel estimation process. Although this allows for a more straightforward (and comparable \cite{fang:24}) discussion of the performance results, still the impact of non-ideal or imperfect \ac{CSI} should be discussed in future works. Nevertheless, note that Algorithm~\ref{alg:MUMISO} could benefit from the channel hardening effect when the direct channel is obstructed. This is so because $\boldsymbol{Y}$ and $\boldsymbol{Z}$ become invariant across different channel realizations for a sufficiently large number of $K$ and $L$, respectively. Consequently, this could simplify the channel estimation process, while making the Algorithm~\ref{alg:MUMISO} more robust to imperfect \ac{CSI}.      

\subsection{Extension to Different \ac{BD-RIS} Architectures}
Throughout this work the fully-connected \ac{BD-RIS} architecture is assumed, for which \ac{BD-RIS} elements are connected to all the remaining elements through adjustable impedances. However, there are other architectures worth mentioning such as the group-connected, tree-connected, and forest-connected, just to name a few. Therefore, adaptations to Algorithm~\ref{alg:MUMISO} considering different \ac{BD-RIS} architectures could be potentially explored. Likewise, aside from the successive diagonal mapping used in this work, other mappings between tree level and \ac{BD-RIS} matrix entries in Algorithm~\ref{alg:MUMISO} could also be further explored. Moreover, the symmetry restriction for the \ac{BD-RIS} matrix could be relaxed, creating new conditions and settings for Algorithm~\ref{alg:MUMISO}.

\commenttxt{\subsection{Extension to \acp{UE} with Multiple Antennas}
A logical extension in the context of this work is considering \acp{UE} with $M$ antennas, that is, \ac{MU-MIMO} systems. This consideration is of practical relevance as multiple antenna systems are becoming increasingly more common in modern communication systems \cite{jema:24}. Since for \ac{MU-MIMO} systems the channel strength maximization problem can be rewritten as \cite{wu:25}
\begin{equation}
    \begin{aligned}
        \max_{\boldsymbol{\Theta}} \ & \sum_{\forall k}{\|\boldsymbol{G}_k^\dagger + \boldsymbol{H}_k^\dagger \boldsymbol{\Theta}\boldsymbol{\Upsilon}\|_F^2} \\
        \text{s.t. } \ & \boldsymbol{\Theta}^\dagger \boldsymbol{\Theta} = \mathbf{I}, \\
                       & \boldsymbol{\Theta} = \boldsymbol{\Theta}^{\top}.
    \end{aligned}
\end{equation}
with $\boldsymbol{G}_k \in \mathbb{C}^{L \times M}$ and $\boldsymbol{H}_k \in \mathbb{C}^{N \times M}$, it follows that line (\ref{line:wUn}) of Algorithm~\ref{alg:MUMISO}, for example, would be modified accordingly as
\begin{equation}
    w \gets \frac{1}{K}\sum_{\forall k}{(\Re\{m^k_l\} + \|\boldsymbol{X}_k\|_*)(2\|\boldsymbol{X}_k\|_*)^{-1}},
\end{equation}
wherein $m^k_l \gets \trace{(\boldsymbol{\Phi}\boldsymbol{X}_k)}$ and, similarly, for lines (\ref{line:wOb}) and (\ref{line:mOb}) in Algorithm~\ref{alg:MUMISO}.

This shows that the proposed tree search algorithm is scalable and readily available to modifications, given its straightforward implementation. Therefore, it would be of interest to evaluate the impacts on Algorithm~\ref{alg:MUMISO} caused by the increase of dimensionality brought by the \ac{MU-MIMO} extension.}

\section{Conclusion}\label{sec:concl}
We have shown in this work that the proposed depth-first tree search algorithm is able to successfully configure the \ac{BD-RIS} by selecting a set of discrete phase-shifts accordingly. Through its heuristic parametrization, the proposed algorithm also demonstrates a remarkable trade-off between channel strength maximization performance and computational complexity scalability. It can be concluded from the numerical results that the proposed algorithm presents a polynomial complexity, while also showing superior performance for the scenario in which the direct channel is obstructed.

Beyond the topics already discussed in the previous section, we expect that this work could motivate further research into search algorithms for \ac{BD-RIS} systems and also serve as a first step towards scalable and accessible configuration solutions. \commenttxt{We also envision that the optimization of the pruning variables, as well as an in-depth sensitivity analysis, could present an interesting opportunity for future research.}  

\bibliography{references}
\bibliographystyle{IEEEtran}

\begin{IEEEbiography}[{\includegraphics[width=1in,height=1.25in,clip,keepaspectratio]{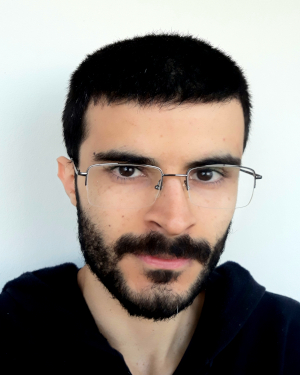}}]{Pedro H. C. de Souza} was born in Santa Rita do Sapuca\'i, Minas Gerais, MG, Brazil in 1992. He received the B.S., M.S. and the doctoral degrees in telecommunications engineering from the National Institute of Telecommunications - INATEL, Santa Rita do Sapuca\'i, in 2015, 2017 and 2022, respectively; is currently working as a postdoctoral researcher in telecommunications engineering at INATEL, under the xGMobile Project. In 2014 he was a Hardware Tester with the INATEL Competence Center - ICC and from 2023 to 2025 he was a postdoctoral researcher supported by FAPESP (\textit{Fundação de Amparo à Pesquisa do Estado de São Paulo}) under the SAMURAI (\textit{Smart 5G Core And Multiran Integration}) Project. His main interests include: digital communication systems, mobile telecommunications systems, 6G, reconfigurable intelligent surfaces, convex optimization for telecommunication systems, compressive sensing/learning, cognitive radio.
\end{IEEEbiography}

\begin{IEEEbiography}[{\includegraphics[width=1in,height=1.25in,clip,keepaspectratio]{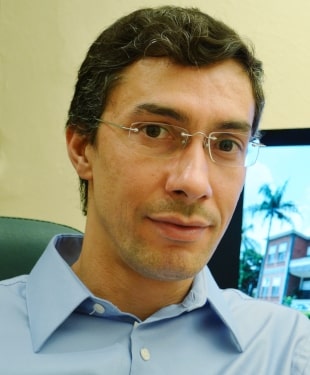}}]{Luciano Leonel Mendes} received the B.Sc. and M.Sc. degrees from Inatel, Brazil, in 2001 and 2003, respectively, and the Doctor degree from Unicamp, Brazil, in 2007, all in electrical engineering. Since 2001, he has been a Professor with Inatel, where he has acted as the Technical Manager of the Hardware Development Laboratory from 2006 to 2012. From 2013 to 2015, he was a Visiting Researcher with the Technical University of Dresden in the Vodafone Chair Mobile Communications Systems, where he has developed his postdoctoral. In 2017, he was elected Research Coordinator of the 5G Brazil Project, an association involving industries, telecom operators, and academia which aims for funding and build an ecosystem toward 5G in Brazil. He is also the technical coordinator of the Brazil 6G Project.
\end{IEEEbiography}

\end{document}